\documentclass[doc]{apa}
\usepackage{graphicx}

\usepackage{amssymb}

 \title{Social Anti-Percolation and Negative Word of Mouth}
 \threeauthors{Tom Erez}{Sarit Moldovan}{Sorin Solomon}
\threeaffiliations{Lagrange Interdisciplinary Laboratory for Excellence in Complexity , ISI, Torino, Italy}{School of Business Administration,
The Hebrew University of Jerusalem, Israel}{Racah Institute of Physics, The Hebrew University of Jerusalem  \\and\\Lagrange Interdisciplinary
Laboratory for Excellence in Complexity, ISI, Torino, Italy} \acknowledgements{corresponding author:\\Tom Erez, email:tom.erez@gmail.com,
\\Fondazione ISI, V. Settimio Severo 65\\Torino 10133, Italy\\The authors thank Jacob Goldenberg and Dietrich Stauffer for their help. This research was supported in part by a grant from the Israeli Academy of Sciences, and by the Yeshaya Horowitz Association through the Center for Complexity Science.}

\abstract{Many new products fail, despite preliminary market surveys having determined considerable potential market share. This effect is too
systematic to be attributed to bad luck. We suggest an explanation by presenting a new percolation theory model for product propagation, where
agents interact over a social network. In our model, agents who do not adopt the product spread negative word of mouth to their neighbors, and
so their neighborhood becomes less susceptible to the product. The result is a dramatic increase in the percolation threshold. When the effect
of negative word of mouth is strong enough, it is shown to block any product from spreading to a significant fraction of the network. So, rather
then being rejected by a large fraction of the agents, the product gets blocked by the rejection of a negligible fraction of the potential
market. The rest of the potential buyers do not adopt the product because they are never exposed to it: the negative word of mouth spread by
initial rejectors suffocates the diffusion by negatively affecting the immediate neighborhood of the propagation front.}

\begin{document}
\maketitle

\section{Introduction}
Many new products fail to meet their expected market share; While preliminary market surveys may report a large portion of potential buyers,
the actual sales might reach only a negligible fraction of the market \cite{BobrowShafer87,McMathForbes98}. Massive scientific research, as well
as large financial, human, media and technological resources, have been invested to improve market sampling. However, there seems to be a
``glass ceiling'' to the success rate of sales prediction. In the present paper we explain this phenomenon in terms of a ``market percolation
phase transition''. We show that the eventual market share of a product depends crucially on the {\bf nature} of interactions between potential
buyers, more so than simply on their number \cite{Bass} or even their network of connections \cite{Solomon00,solweis99,born}.

\subsection{Percolation Theory}
In general, percolation theory describes the emergence of connected clusters. Historically, percolation problems were first studied in
chemistry, when Flory and Stockmayer studied Gelation as a percolation process on a Bethe lattice \cite{gel}. Since then, percolation theory has
been developed extensively by both mathematicians and physicists \cite{stauffer}, and was applied to a variety of other subjects, from
epidemiology to oil fields and forest fires \cite{havlin}.

The main phenomenon studied in percolation models is the emergence of a phase transition: a dramatic change in the qualitative behavior of the
system, triggered by an infinitesimal change in the the parameters. For example, a small difference in virulence can make the difference between
a seasonal flu and a global epidemic. Percolation theory offers a broad body of knowledge for the study of such phenomena, and by casting a
problem in percolation terms one can gain access to a wide set intuitions and rigorous results.

In short, percolation models involve agents that interact across a network. The interaction consists usually of influencing the state of a
neighbor agent (i.e. a ``sick'' agent can potentially change the state of its ``healthy'' neighbors by infecting them). Obviously, the affected
neighbor can now further affect one of its neighbors, and so allow the effect to diffuse across the network. Percolation theory studies the
conditions in which the set of affected agents reaches a macroscopic size (a non-vanishing fraction of the entire set of susceptible agents).
Interestingly enough, Percolation theory predicts that often such a ``global'' diffusion will not take place, as the propagation may die out
before any significant fraction of the system is reached by the diffusion dynamics. The transition to the percolating regime, where almost all
susceptible agents are infected, is usually very sharp. The values of the parameters at which this happens are called ``critical values''. As
one varies the parameters through their critical values, the system abruptly passes from the ``seasonal flu'' phase to the ``epidemic'' one.
This is the famous percolation phase transition.

Mort \citeyear{Mort91} suggested the application of percolation theory to marketing: a product spreads among adopters, and can be said to
percolate (or not) through the social network. Solomon and Weisbuch \citeyear{solweis99} proposed to look at ``Social Percolation'', and regard
society as a network through which a social phenomenon (information / belief / product / behavior) may, or may not, percolate. In the right
conditions a macroscopic cluster of adopters emerges, and most of the susceptible people will be eventually influenced. However, if the adoption
rate (``social virulence'') or the typical number of neighbors per agent are below their critical values (``percolation threshold'') the spread
would stop before any significant fraction of the susceptible adopters is reached. As seen in Figure 1, if the proportion of susceptible agents
is below the percolation threshold, they form disjoint little islands (Fig. 1a). In this case a propagation that starts on one of the islands
can ``conquer'' this entire island, but cannot ``spill'' beyond its borders. However, as the proportion of susceptible agents increases,
eventually bridges will form between a significant fraction of the islands. This is  the percolation Transition: the clusters expand to where
they touch each other, and form one giant percolating cluster (Fig. 1b).

It is not simple to introduce percolation models to marketing science, mostly because percolation theory deals with ``micro'' data, on a
fine-grained scale, and so one needs spatial micro-data in order to tune the model's parameters. For social networks, such information on the
micro-structure is usually unknown; spatial micro-distribution of sales data is usually not publicly available. However, the approach has gained
support in some recent studies - for example, Rosen \citeyear{rosen} reports that when Hotmail email service was launched it was quickly spread
through word of mouth, and although email is not a ``local'' service, it was still spreading in local areas. In addition, scientists working in
percolation theory are usually not familiar with marketing science, its basic questions or its methods. State-of-the-art simulations in
percolation theory may look like video games to a marketing scientist, while for a physicist who is not familiar with the particularities of
marketing literature, the quantitative message might look unrecoverably buried in particular details specific to the case under study.

In this chapter we extend the standard percolation model by introducing effects of negative word of mouth (NWOM). The resulting model was never
considered before by percolation theory, probably because in the domains usually studied by it there are no negative effects between neighboring
agents. We do not claim that the simplified model we present can be directly applied to a marketing scenario - this would require calibrating
the model against data on the spatial and temporal distribution of sales (which the companies are very reluctant to release). However, since the
main dynamical implications of the model are quite dramatic (large jumps in the sales for small parameter changes) they are likely to be
replicated in a wide range of real systems.  The present model should be considered as setting the scene for a tighter interdisciplinary effort
by providing a formal model for a crucial market force : the Negative Word of Mouth.

\subsection{Negative Word-of-Mouth in Marketing}

From marketing research we learn that NWOM has a profound effect on adoption patterns. For example, Leonard-Barton \citeyear{LB85} found that an
innovation reached only 70\% of its market potential due to resisting consumers. In addition, Herr et al. \citeyear{Herr91} found that NWOM may
decrease product evaluations. Two main characteristics of NWOM have been stated in marketing literature: it is more informative than positive
word of mouth, and thus may have a stronger effect \cite{HauserUrbanWeinberg93,Herr91}, and it may be contagious and spread independently of
exposure to the product \cite{MarquisFiliatrault02}. Both these attributes are addressed by our model.

Some models incorporating NWOM were suggested in marketing theory (e.g.
\cite{KalishLilien86,MahajanMullerKerin84,Midgely76,MoldovanGoldenberginpress}), but none of these models address the context of the underlying
social network. The standard way to describe market share progression in marketing theory is by using differential equations, and most
models used today are extensions of the classical Bass model \cite{Bass}. Predictions on adoption dynamics are made by fitting the parameters of
these differential equations according to empirical sales data. The main criticism on the ODE approach is that it is inefficient at the early
stages of product spread, as it is too sensitive to early fluctuations in the adoption process \cite{newzealand}. This renders the forecast
method less efficient, because by the time enough data points were collected, most of the ``damage'' (e.g. over- or under-production) is already
done. We believe that this failure is inherently linked to the fact that the Bass framework pre-assumes ``mean-field'' conditions (i.e. all
agents may interact with all others), and ignores the underlying social network.

Marketing theory distinguishes between two types of forces affecting the consumer - the ``internal'' (local) force, a name for all the
influences cast on individual consumers by peer consumers (e.g. word of mouth), and the ``external'' (global) force, the kind of influences that
are cast upon all  consumers equally  (e.g. marketing efforts). It is accepted in marketing that on the long run, internal forces are of greater
significance for the adoption pattern of a product (although the external force is important for product awareness, or for ``activation'' of the
internal force). An estimation of the effect of the two forces, internal and external, shows that after takeoff, the word of mouth effect is 10
times larger than marketing efforts \cite{GoldenbergLibaiMuller01} and may be responsible for as much as 80\% of the sales \cite{Mahajan90}.

Despite the fact that the internal force is more powerful than the external force, the standard effector employed for marketing is advertising,
often by addressing an entire market. This introduces a bias into marketing research as well, which tends to focus on the external forces. The
Social Percolation (SP) framework models internal forces as local interactions between neighboring agents on a social network, and observes the
properties of the resulting adoption patterns. Investigating the effects of internal forces in this manner can provide useful insights to market
behavior, even if we ignore overall effects of external forces (the framework of social percolation can easily be extended to incorporate
external forces as well\cite{globalschemes}). In this chapter we divide the internal force into positive and negative effects, and explore the
effect of negative word of mouth in the SP framework. This provides a focus on the internal force in a structured context.

\section{The Model}
\subsection{Classical Social Percolation}

We start by repeating the ideas of the SP framework \cite{Solomon00,solweis99,born}. The basic element of the model is an agent, representing a
consumer. Each agent $i$ is characterized by a ``preference'' value, $p_i$, and agents are placed in a social network with a fixed structure.
 The model is made complete by introducing a product, represented by a global ``product quality''
value, denoted $Q$. The product spreads in the network between neighbors - whenever an agent adopts the product, it exposes its neighbors to the
$Q$, and they decide individually whether to adopt it or not. Simulating this dynamics is done by exposing a ``seed'' group of agents to the
product, and allowing it to spread from them until its diffusion naturally ends, either because it reaches the network's boundary, or because it
is blocked by non-adopters. Transmission of the product between neighbors happens according to the following adoption rule:

\begin{quote}ADOPTION - Agent $i$ will adopt the product if a neighboring agent adopted the product AND the product's quality is higher than
the agent's preference: $Q>p_i$. \end{quote}

Thus, the potential market for a product consists of all the agents whose $p$ is smaller than $Q$, but as long as this potential market forms
only disconnected clusters, the product cannot reach all ``islands'', and its diffusion is limited.

Although the parametrization of the model with a fixed $Q$ suggests a false interpretation that a unique well-defined value of quality may be
assigned to a product, we do not claim this to be the case. Even as different consumers may have different perceptions of a product's quality,
we choose to incorporate all such inter-personal diversity in the variability of the random values $p_i$. Since adoption is determined by
comparing the ``global'' $Q$ to the ``private'' $p_i$, introducing variance to $Q$ over the agents is equivalent to introducing variance to the
agents' $p_i$.

\subsection{Incorporating Resistance into Social Percolation}
Now we introduce our variant of the model, including NWOM. In what we described so far, the case $Q<p$ had no consequences: failing to meet a
consumer's standards only meant that the product was ignored and the information about the product was not passed on to the consumer's
neighbors. In the present model, we see this case as the equivalent of ``disappointment'', one of the roots of NWOM.

In order to introduce NWOM to the model, we propose to look at another property of the product, uncorrelated with its ``value'' (that is
represented by $Q$). We wish to refer to the product's ``susceptibility to NWOM'', and parameterize it as a number between $0$ and $1$,
following the (trivial) assumption that certain products are more sensitive to NWOM than others. The value chosen may be related to the
marketing strategy, the degree of novelty, and to many other aspects of the product domain. This parameter pertains to the response created in a
potential customer after an encounter with the product that results in non-adoption: If agent $i$ was exposed to the product, but has not
adopted it ($Q<p_i$), we may say that the agent``resisted'' the product. In that case, the agent may spread resistance to the product over her
neighbors, generating NWOM. The effect of such negative spread is that the affected neighbors become less receptive towards the product, i.e.
their chances of adopting it decrease. We model this effect in the following way: in the case of $Q<p_i$, we denote $p_i-Q$ as $D_i$ (note that
$D_i>0$ always). We introduce a parameter $a$, denoting the extent one agent's resistance influences its neighbors. The spread of resistance
(i.e. NWOM) is modeled by increasing $p_j$ by $aD_i$ for all agents $j$ that are neighbors of $i$.

This increase of $p_j$ will have no effect if $j$ has already adopted the product. Yet, if $p_j<Q$ and $p_j+aD_i>Q$, then this agent is said to
have been blocked by NWOM. The increase of $p_j$ is additive - if several agents ``project'' resistance on one agent, all their individual NWOM
effects add together on top of the original $p_j$. So, even if an agent was blocked, subsequent NWOM events will still make things worse for the
product, because the NWOM this agent is prone to spread will be bigger (because $p_j$, and consequently $D_j$, will be bigger); Also, if
$p_j+aD_i<Q$, this change has no immediate effect (as $j$ may still adopt if one of her neighbors will expose her to the product), but this
agent is now more prone to blocking by subsequent NWOM.

Since $Q$ is limited to the range [0,1], an agent $i$ with $p_i>1$ will reject all products, the same as if $p_i$ were $1$. Yet, since $p_i$ is
allowed to grow freely beyond one, the resistance ($D_i$) that this agent may cast on its surrounding is essentially unbound. This allows the
spread of NWOM to ``block'' completely the affected agents. In addition, we chose to model NWOM spread as occurring on a faster time scale than
exposure to the product. This means that in case of resistance spread, the increase of all $p_j$ happens instantly, before any further exposures
of new agents to the product are considered. The rationale is simple: the typical time scale for casting NWOM is one conversation with one
friend; Exposure to the product, on the other hand, is a slower process - the potential customer has to act (e.g. visit the point-of-sale) in
order to potentially acquire the product (in a sense, one can say that ``bad news travel faster'').

In accordance with Marquis \& Filiatrault \citeyear{MarquisFiliatrault02}, we characterize another parameter, $b$, that pertains to the
product's susceptibility to ``bad rumors'' - NWOM that is not based on actual exposure to the product. Hence, the effect of resistance may
travel to second-neighbors as well, and their $p$ is increased by $bD_i$. At first glance, this modeling scheme seems primitive - why not extend
this further, and parameterize the effect to the $n$-th neighbor? Yet we wanted to avoid ``giving wings'' to such rumors - if NWOM is allowed to
propagate freely on the social network, an unrealistic dynamics occur as the NWOM behaves like another product. Therefore we restrict our
attention to spread of NWOM up to second-order neighbors only. Although it may be argued that $a$ and $b$ are dynamic properties, and bear
inter-personal variance, in our simplified abstract model they are taken as a static scalar parameter, depending only on the product.

In summary, we introduce the following rule of interaction to the SP framework, in addition to ADOPTION:

\begin{quote}RESISTANCE - If $i$ was exposed to the product and $p_i>Q$, then \\ for every agent $j$ neighboring $i$,
$p_j$ is updated to $p_j+aD_i$\\
AND\\
for every agent $k$ that is a second neighbor of $i$, $p_k$ is updated to $p_k+bD_i$\end{quote} Thus, every case of adoption generates potential
for further adoption, and every case of disappointment casts a ``cloud'' of NWOM around it. Figure 2 summarizes all rules of interaction.

\section{Simulations}
\subsection{Method}

The indexed set $\{p_i\}_{i=1}^N$ that specifies the ``personal'' values of $p$ for every agent was randomly generated at the beginning of every
simulation, with every such $p_i$ chosen uniformly from the range [0,1]. At every iteration, values are fixed for $Q, a$ and $b$, and the
simulated dynamics start from a ``seed'' of one adopting agent selected at random, who spreads the product to her neighbors. These agents are
added to the ``adoption front'', a list of agents who are waiting for evaluation of the product. At every time step, an agent $i$ from the
adoption front is randomly selected, and her current value of $p_i$ (possibly increased by previous spreads of resistance) is compared to $Q$.
If $Q>p_i$, the agent is marked as 'adopter', and all her neighbors are put on the adoption front, in line for exposure to the product. If
$Q<p_i$, $aD_i$ is spread to $i$'s first neighbors and immediately added to their $p_j$, and $bD_i$ is spread to $i$'s second neighbors,
immediately added to their $p_k$. Thus, the front expands away from the starting point (seed) in a random fashion, and the iteration ends when
the front exhausts itself. This happens either when it traverses the entire lattice and reaches the boundaries, or when all the agents on the
front rejected the product (so their neighbors were not added to the front). The first case corresponds to the product meeting its market
potential, passing the percolation threshold and traversing the network, and the second corresponds to the product being blocked by the
population of agents.

\subsection{Percolation Threshold Measurement}

The product's ``strength'' may be measured in two ways: a percolation measure - whether the product percolated successfully through the lattice
`from side to side'; and market penetration - what is the size of the adoption cluster. Notice that these two measurements may not coincide, as
a product may percolate and still achieve a small adoption rate. In regular lattices of three dimensions and up this happens near the
percolation threshold, where the percolating cluster has a fractal structure and a minimal density.

In the present work we focused on measuring the percolation threshold. in order to determine whether a product 'percolated', we checked how far
it spread across the network, in terms of the distance from the initial seed. We looked at the ``shell'' structure of the network around the
initial node, marking all the nodes found at distance $d$ from the initial seed as belonging to shell $d$. Then we identified the biggest shell
and percolation was marked successful if at least one agent from the biggest shell adopted the product. This definition is equivalent with
classical measures of percolation on regular lattices, and it generalizes it to cases of irregular networks as well.

In the current chapter we focus our attention to a regular lattice with non-periodic boundary conditions; we consider it an idealized
``zero-order'' approximation for a social network. Through this simplifying assumption we render our model less realistic, but more accessible
to theoretical analysis. However, the model's definition is not dependent on the topology of the network, and the same model can be run on
networks with arbitrary topologies.

We employed a binary search method of dichotomy for the estimation of the percolation threshold. For every instance of random values $\{p_i\}$,
the threshold $Q_c$ is a number such that for $Q>Q_c$ the product percolates, and for $Q<Q_c$ the product does not percolate. We estimated it by
repeatedly exposing the lattice to different $Q$'s, and changing $Q$ in an adaptive way, according to the success or failure of the last
iteration. If at iteration $i$ the product of quality $Q_i$ percolated, in the next iteration $Q$ will be increased:
$Q_{i+1}=Q_i+{\frac{1}{2^{i+1}}}$; accordingly, if at iteration $i$ the product was blocked, $Q_{i+1}$ will be decreased by the same amount.

Due to resistance spread, the values of $\{p_i\}$ change in the course of an iteration, and so we reset them to their original values between
iterations. For every instance of random values $\{p_i\}$ and fixed $a$ and $b$, we did 10 iterations of adaptive $Q$, and so effectively
estimated the threshold within an uncertainty margin of size ${\frac{1}{2^{10}}}$.

\section{Results}

In the present work we measured how negative word-of-mouth affects the percolation threshold. This was simulated by applying different values to
$a$ and $b$, and measuring the percolation threshold under these conditions. Figure 3 shows the threshold's dependence on $a$ and $b$ for lattices
of various dimensions.

At the corner of $a=b=0$, the measured threshold corresponds to the classical percolation threshold for these topologies \cite{stauffer}. As $a$
and $b$ increase, the threshold increases, until it saturates near 1 for high enough values. The increase of both $a$ and $b$ causes an increase
of the threshold; however, it is clear to see that the dominant effect is that of $b$, with an effective threshold of $>0.9$ for $b$ as low as
$0.55$ in the 4D case (Fig. 3).

Of course, the number of second neighbors increases with lattice dimension (8 in 2D, 18 in 3D, and 32 in 4D), but on a careful observation, a
stronger, autocatalytic effect can be noticed, which is a result of the synergy between the neighbors. Since the front spreads from one point
out, two nearby agents have big probability to be on the front at the same time. Since resistance spreads from agents who are on the front,
their neighbors, hit by the local effect of NWOM, are probably on the front as well. These are exactly the agents that will be evaluated next,
and therefore an increase to their $p$ is the most significant, because they will certainly be exposed to the product and may spread NWOM (see
Fig. 2c-d). In a sense, the NWOM hits the product at the most sensitive group of consumers - if the resistance would be spread to far-away
agents, the front might only get there in a long time, or never get to that part of the network, which will render the blocking impact of the
resistance less effective. Without any long-range effects, the local spread of resistance ensures maximum negative impact to the NWOM, while
affecting only a finite number of agents (and so their proportion approaches $0\%$ in the limit of large networks).

In the standard percolation model, the threshold decreases as the dimensionality of the lattice increases. This is due to the increase in the
average number of neighbors, which brings about an abundance of paths between every two nodes. This number increases with the dimension, and
therefore the product is less likely to be blocked, because there is greater probability of finding at least one path through which the product
can percolate. At the extreme case of a highly connected irregular network, for example with a power-law distribution of the degrees, the
percolation threshold was shown to be exactly 0 \cite{barabasi}. However, when NWOM is introduced, the increased connectivity of the network has
the opposite effect - since resistance can spread to more neighbors, it is prone to have a greater impact on the front, and hence on the
progress of the product. Since resistance is spread only from agents who are evaluated, i.e. on the front, the impact of NWOM on the front is
greater when more agents on the front are close to each other. This number increases with the dimension - in 2D, every one of the 8 second
neighbors has 4 of the other as her own second neighbor; in 4D, each one of the 32 second neighbors has 12 of the others as second neighbors.

It is worth noting that all the above regular lattices are without triangles. Since only first neighbors are added to the front in case of
adoption, every group of agents being added to the front has no links between its members. A more interconnected front is more susceptible to
effects of $a$, because every event of resistance would spread NWOM to more agents on the front. In a network with a greater clustering
coefficient, $a$ is prone to have a bigger effect. To demonstrate this, we compare between a regular 2D lattice with a 4-neighbor topology
(von-Neumann) and 8-neighbor topology (Moore). The results are presented in figure 4, where it can be seen that the increase of the percolation
threshold due to the increase of $a$ in the 8-neighbor lattice is greater.

Another result is the emergence of local resistance leaders. In the classic percolation model, cluster size increases monotonically with $Q$.
This is not the case when NWOM is incorporated into the model: for a given randomization, the cluster size may decrease with increasing $Q$
(Fig. 5). This surprising result of the NWOM dynamics is due to particular micro-level setting: if in a particular section of the network the
first agents exposed to the product are of high $p$, they are bound to spread significant resistance, and perhaps block the entire section. We
call such agents ``Resistance Leaders'', as it is certain that they would be resistant to most products.

The blocking effect of such resistance leaders depends very much on the particular circumstances of the product spread - the question whether a
resistance leader, or rather her neighbor, would be exposed first to the product might have a crucial effect on the progress of the product's
spread in that section, and this depends totally on the micro-conditions there. If most of the agents in her neighborhood already adopted the
product, the NWOM spread by an resistance leader may have little effect. This novel feature of the model happens both above and below the
percolation threshold.

\section{Discussion}

As for the value of the current model per-se to marketing science, we can propose the following perspective: if a product is advertised,
different parts of the lattice may be exposed to it, and the chances of a product being completely blocked are lower (although exist, see for
example \cite{gold2005}). The present model might be more directly applicable for products and services that are hardly advertised, but instead
spread through word of mouth (e.g. restaurants, shops, high risk and radical innovations).

The model we presented here introduces two novel parameters to the percolation model - the negative impact of non-adoption on the first
neighbors ($a$) and on the second neighbors ($b$). These parameters are not a vague idealization - their real-world counterparts are easy to
define: the parameter $a$ is related to the nature of the disappointment caused by an event of non-adoption. Consumers can be disappointed and
reject an innovation due to several reasons, such as a publicity campaign that creates high expectations that are not confirmed \cite{complain},
fear of the new and unfamiliar, or resistance to change \cite{ashesh,futurist}, low expectations from a new technology, no benefits or high
pricing \cite{fads,complain}. The main point is that disappointed consumers tend to spread more NWOM and have higher effect on other consumers
\cite{Herr91,Richins}.

Moreover, consumers may spread NWOM to their friends just on the basis of their exposure to negative information, and without any trial or
contact with the product. The parameter $b$ represents the extent to which consumers tend to tell each other stories about products they never
tried. The above reasons for rejecting the innovation can serve as reasons to spread NWOM further on, a phenomena that was found in previous
studies \cite{MarquisFiliatrault02}. Leonard-Barton \citeyear{LB85} found that 20\% of dentists were familiar with, yet rejected, a $successful$
dental innovation; many of them were not even willing to try it as a result of NWOM. Consistent with the theme of ``bad news travels faster'',
consider the following anecdotal report of a major taxi company in New Zealand which lost almost 60\% of its business as a result of an angry
customer spreading her story to thousands of women throughout New Zealand \cite{venganve}.

These cases can be documented, but can hardly be understood or predicted in the classical (Bass-like) framework. Standard techniques of
estimation of the potential market give little or no attention to the emergent effects of consumer interaction. Focus groups and random sampling
may give an accurate estimate of the potential in a naive market, but as soon as the product is actually introduced, the naive market changes
shape by effects of word of mouth whose source are the early adopters. A product may seem to be good enough at the preliminary probing of the
market, and yet fail due to the devastating effect of NWOM.

Therefore, it would be very helpful if marketing research could estimate the parameters of the endogenous social interaction of the market. A
more complexity-aware probing of the market should also estimate the ``interaction value'' of a disappointed customer, paying attention to the
probability of people discussing the product without ever being exposed to it. Our model shows that these aspects of the product and the market
context have great impact on the eventual success or failure of the product.

Since this novel approach challenges the traditional methods for sales forecast, we do not expect it to pass without resistance. Nonetheless,
the scientific method revolves around falsifiability - since there is a radical difference between the theoretical underpinning of social
percolation and aggregate, Bass-like models, one should seek experimental settings where distinct predictions can be offered by the rival
approaches. Such tests are the only reliable way to discern between the two theoretical approaches and transcend the bias inherited by
disciplinary tradition. The main prediction of the present anti-percolation framework is that in the case of a failed product, many of the
potential adopters will never feel the wave of propagation, and are not at all acquainted with the product. This is at variance with the
predictions of aggregate models, where the the entire population is either adopting or explicitly rejecting the product. Thus, it could be quite
straightforward to discriminate empirically between the various models predictions.

In summary, we believe that percolation theory (with the present anti-percolation amendments) has much to contribute to marketing research. We
hope that the marketing world will recognize the potential in this mature and rich theory, embrace the language of agent-based models and
simulations, and focus its attention on designing and testing the predictions of increasingly realistic models.

\bibliography{natinsp}
\newpage
\begin{figure}[tp]
\includegraphics[angle=0,scale=0.7]{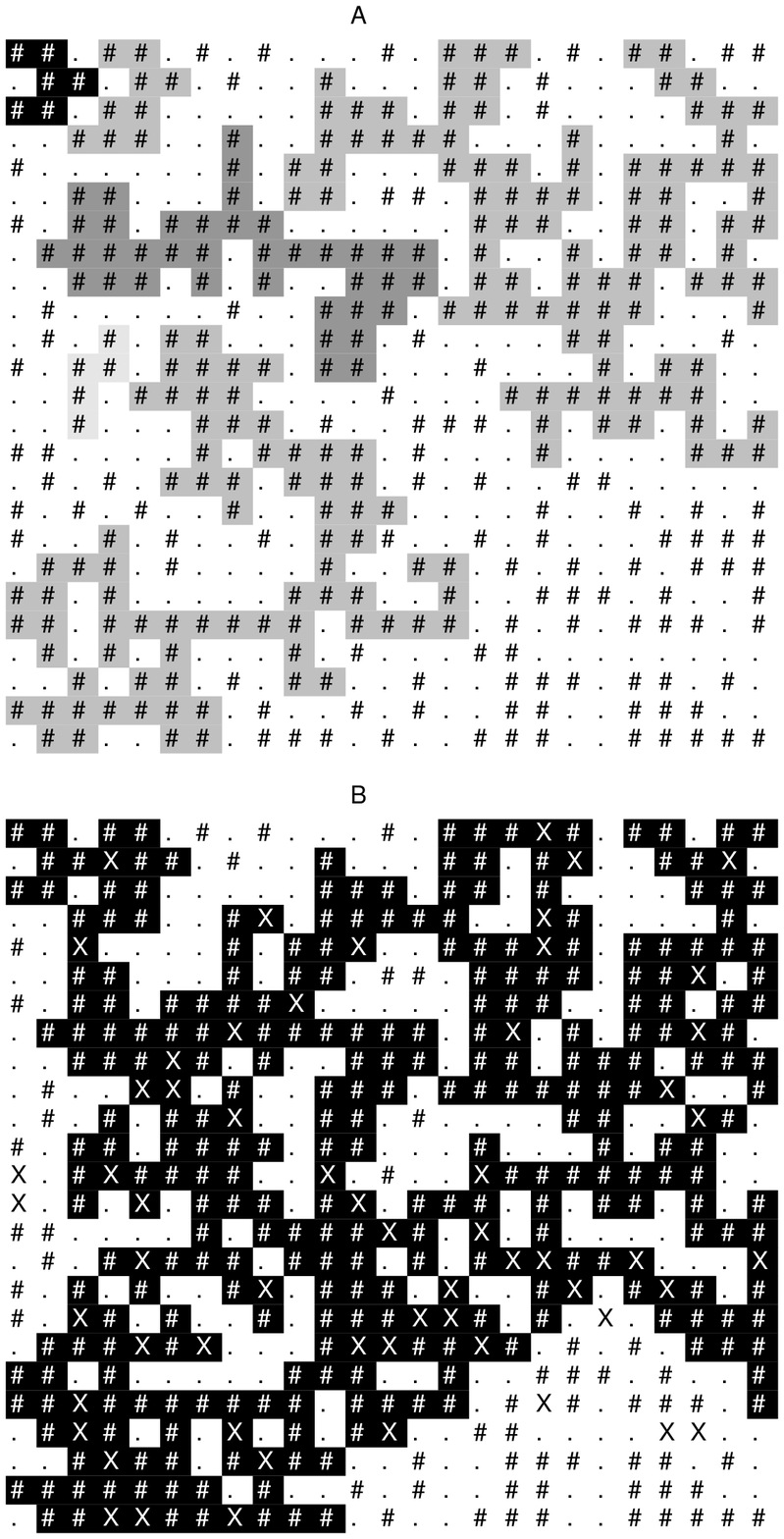}
\caption{(a) A 25x25 regular grid in which a half of the cells were randomly marked with \#. The resulting cluster structure is one of several
disconnected islands. If a propagation starts from the upper left corner, it will not diffuse beyond the cluster marked in black. The
propagation cannot reach other clusters, some of them marked in grey. (b) if an additional 10\% is marked with X, the cluster structure changes
dramatically, and a cluster proportional to the entire grid size emerges. Now a propagation from the upper right corner can span the entire
network. (Neighborhood links are drawn along the main axes, so every agent has 4 neighbors at most: N, S, E and W.)}\label{fig:fig1}
\end{figure}

\newpage
\begin{figure}[tp]
\includegraphics[angle=-90,scale=0.5]{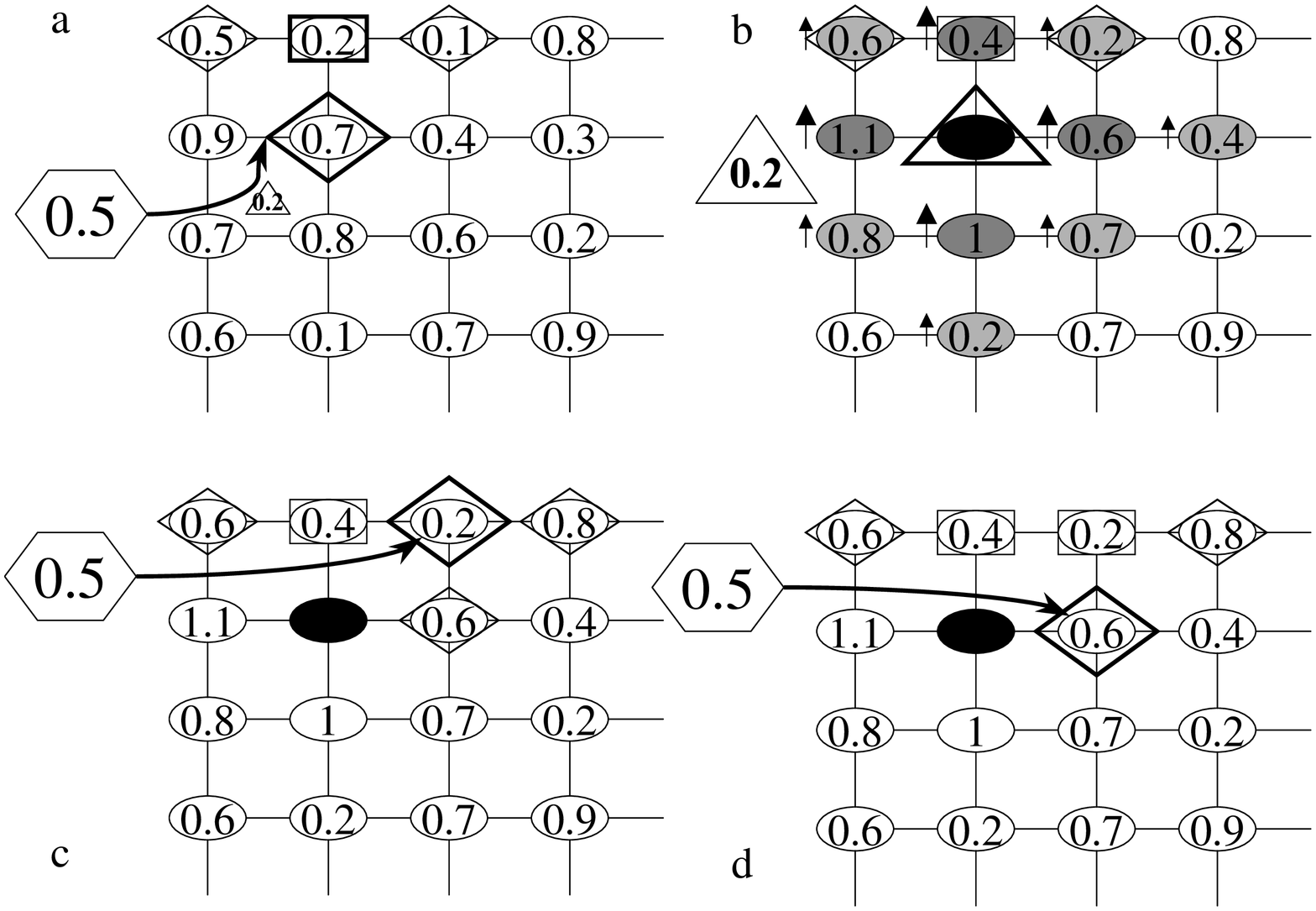}
\caption{(a) $Q$, marked as a hexagon, is $0.5$. The agent marked in square is the seed, with $p=0.2$, and so it adopts the product and all its
neighbors are added to the front (marked in diamonds). Next, the agent marked in a bold diamond is evaluated. Since its $p=0.7$ is bigger than
$Q$, this agent dos not adopt, but rather spreads resistance ($D=0.2$), marked in a triangle. (b) The agent is now marked in black, indicating
that it will not adopt anymore. The resistance is immediately spread from the agent in a triangle to all its neighbors: for $a=1$ and $b=0.5$,
the first neighbors' $p$ increases by $0.2$ (marked in dark grey), and the second neighbors' $p$ increases by $0.1$. Red arrows to the left of
the agents mark the increase of $p$. Note how the agent to its left, whose $p$ was $0.4$, now has $p=0.6$. This agent is no longer a potential
buyer, but a potential spread of resistance; it was blocked by the NWOM. (c) Next, another agent is selected from the front. This agent is now
marked in a bold diamond, and since its $p<Q$, it adopts the product and all its neighbors are added to the front. (d) Another agent is selected
from the front. This time its $p>Q$, and this will cause spread of resistance. This agent was originally part of the potential market, but was
blocked by NWOM, and now will generate more NWOM. This illustrates the auto-catalytic nature of NWOM.}\label{fig:fig2}
\end{figure}

\newpage

\begin{figure}[tp]
\includegraphics[scale=0.5]{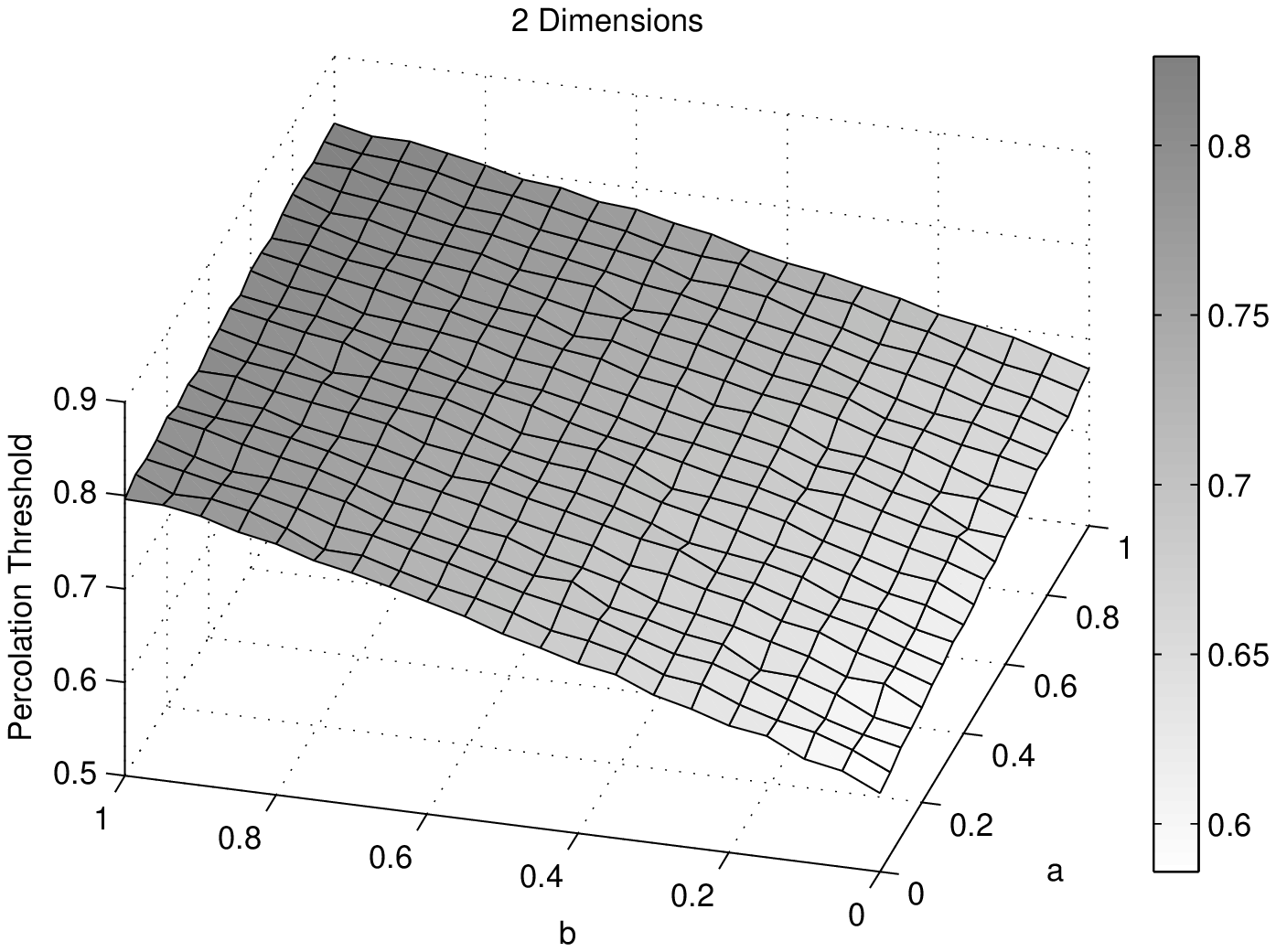} \\  \includegraphics[scale=0.5]{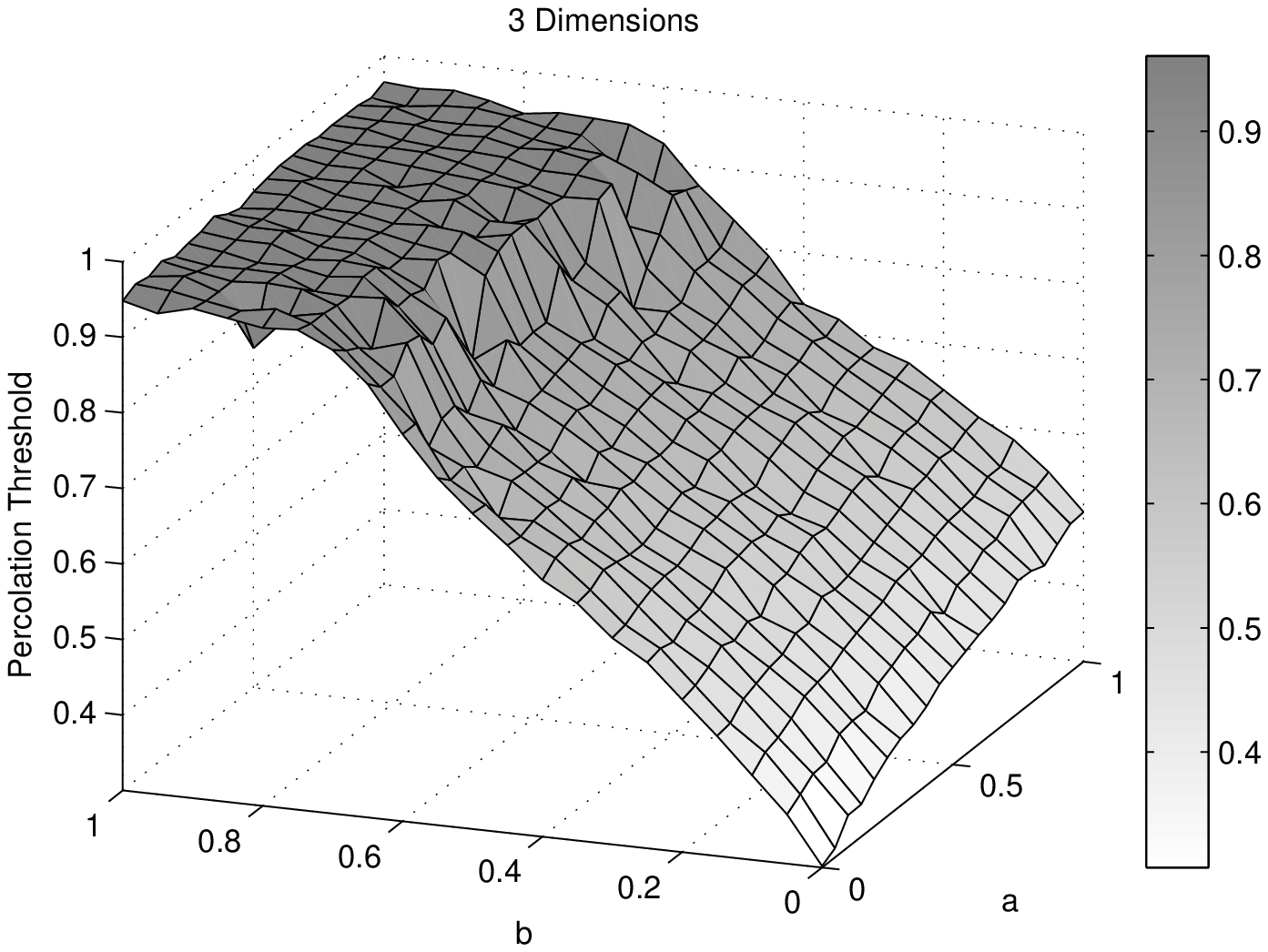} \includegraphics[scale=0.5]{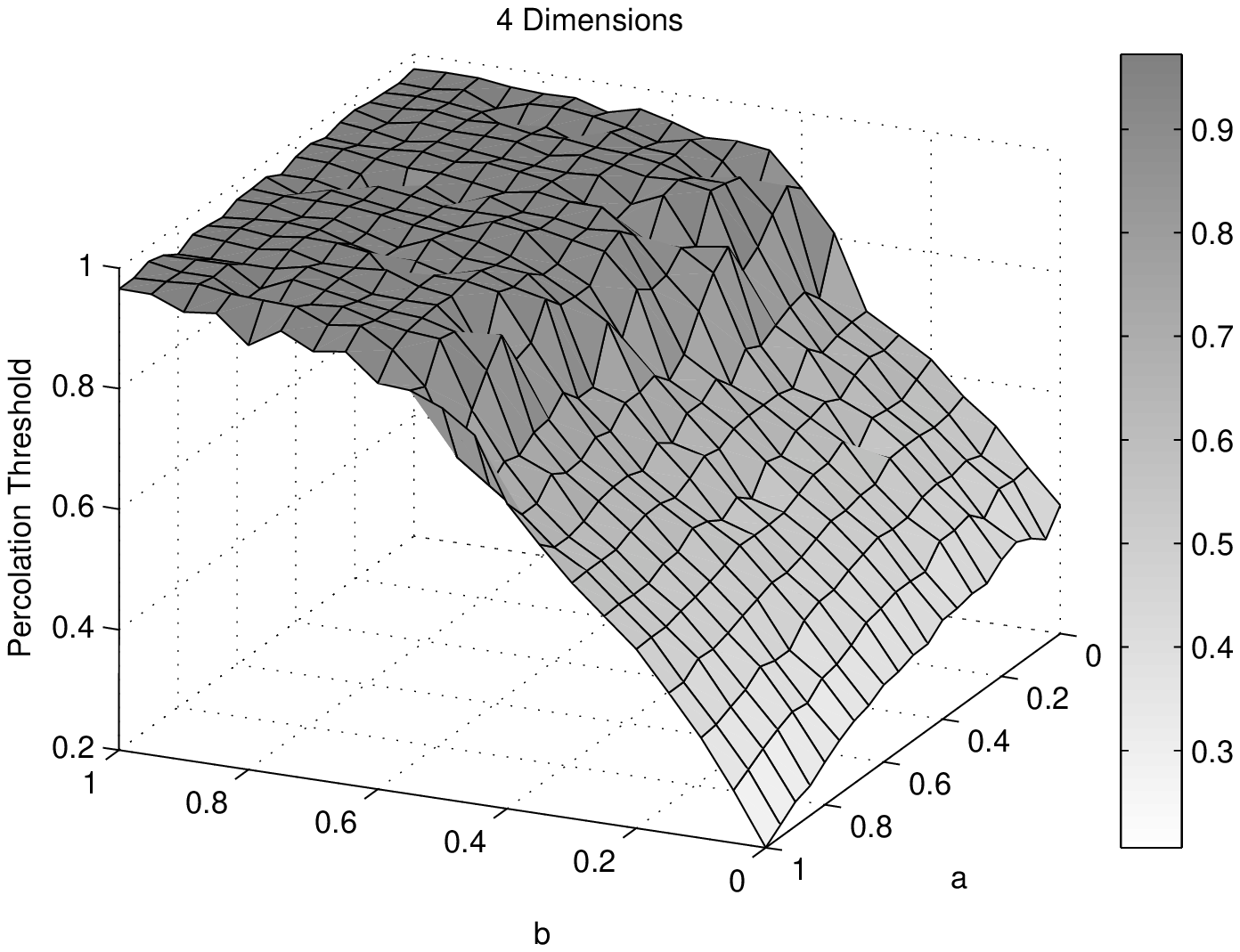}
\caption{These plots show the the percolation threshold (Z axis) as dependent on the parameters $a$ and $b$ (the X-Y plane). In the corner of
$a=b=0$ the percolation threshold corresponds to the standard results, and it increases with both $a$ and $b$. However, it is evident that $b$
has a greater influence, and above a certain level, the network becomes totally impenetrable to the product. Figure 2a is for 2
dimensions:$1000^2=1,000,000$ agents; Figure 2b is for 3 dimensions: $80^3$=512,000 agents; Figure 2c is for 4 dimensions:$40^4$=2,560,000
agents.} \label{fig:fig3}
\end{figure}

\begin{figure}[tp]
\includegraphics[angle=-90,scale=0.5]{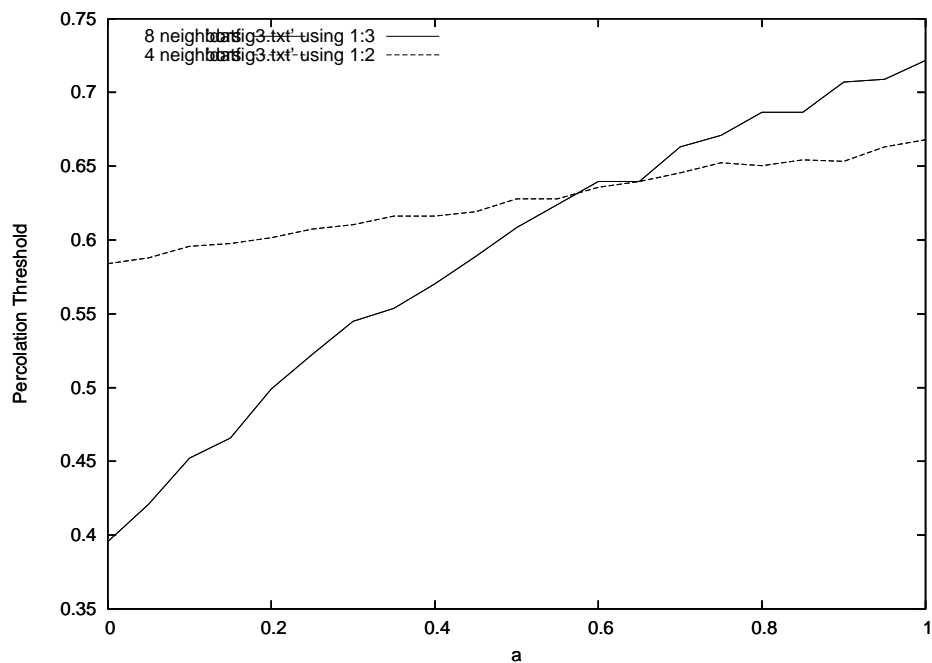}
\caption{Here the importance of the clustering coefficient is demonstrated by comparing the effect of $a$ on two different two-dimensional
topologies: 4 neighbors (dashed line) and 8 neighbors (full line). In the 4 neighbor topology the front is not a disconnected set, since newly
added agents would have links with other agents on the front already; but the front is more interconnected in networks of greater clustering
coefficient. Thus, the network is more susceptible to effects of $a$, because more agents on the front would be affected by every event of
resistance.} \label{fig:fig4}
\end{figure}
\begin{figure}[tp]
\includegraphics[angle=-90,scale=0.5]{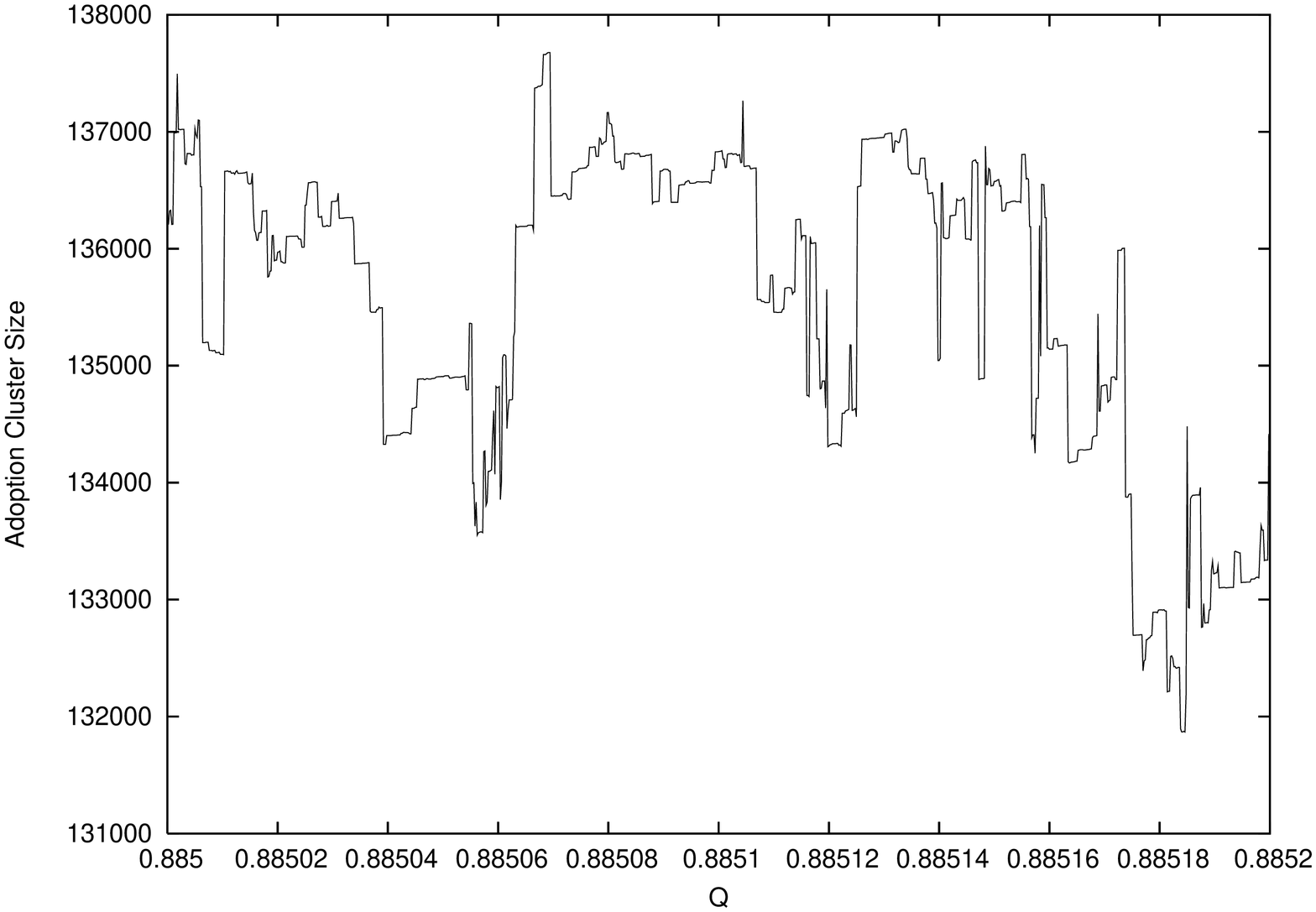} \includegraphics[angle=-90,scale=0.5]{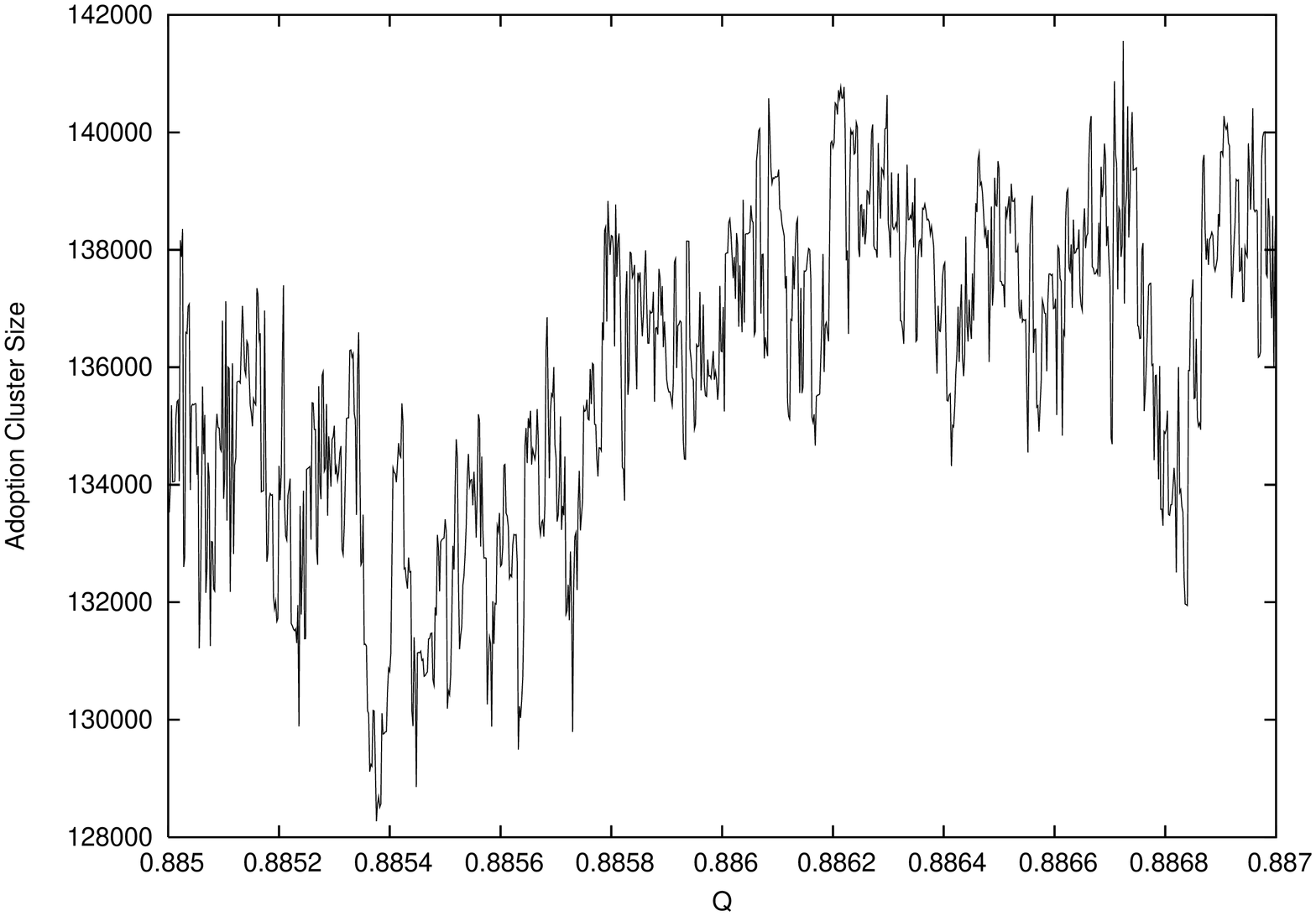}
\caption{The dependence of the percolation on $Q$ is non-monotonic, and a better product may have smaller sales; this is due to the effect of
local resistance leaders, agents with high $p$ that block their entire neighborhood if the product reaches them first.} \label{fig:fig5}
\end{figure}

\end{document}